\documentclass[11pt]{article}
\usepackage{sigsam, amsmath}
\usepackage{amsthm}
\usepackage{amssymb}
\usepackage{amsfonts}
\usepackage{mathtools}
\usepackage{maple}
\usepackage[utf8]{inputenc}
\usepackage{breqn}
\usepackage{listings}
\usepackage{moreverb}
\usepackage{verbatim}
\usepackage{sverb}
\usepackage{hyperref}
\usepackage{xcolor}
\usepackage{url}
\usepackage{cleveref}
\newtheorem{theorem}{Theorem}

\newtheorem{example}[theorem]{Example}
\providecommand{\keywords}[1]{\textbf{\textit{Keywords:}} #1}
\newcommand*{\QEDA}{\hfill\ensuremath{\blacksquare}}

\newcommand{\NN}{\mathbb{N}}

\newcommand{\KK}{\mathbb{K}}

\issue{ISSAC'23}
\articlehead{Software Demonstration} 
\titlehead{Operations for D-Algebraic Functions}
\authorhead{B. Teguia Tabuguia}
\begin{document}
	
	\title{Operations for D-Algebraic Functions}
	
	\author{Bertrand Teguia Tabuguia\\
		Nonlinear Algebra Group\\
		Max Planck Institute for Mathematics in the Sciences/ MPI for Software Systems\\
		04103 Leipzig/ 66123 Saarbrücken, Germany\\
		{\tt bertrand.teguia@mis.mpg.de}}
	
	\date{}
	
	\maketitle
	
	\begin{abstract}
		A function is differentially algebraic (or simply D-algebraic) if there is a polynomial relationship between some of its derivatives and the indeterminate variable. Many functions in the sciences, such as Mathieu functions, the Weierstrass elliptic functions, and holonomic or D-finite functions are D-algebraic. These functions form a field, and are closed under composition, taking functional inverse, and derivation. We present implementation for each underlying operation. We also give a systematic way for computing an algebraic differential equation from a linear differential equation with D-finite function coefficients.
		
		Each command is a feature of our Maple package \texttt{NLDE} available at \url{https://mathrepo.mis.mpg.de/DAlgebraicFunctions/NLDEpackage}.
	\end{abstract}
	
	
	\maketitle
	
	\section{Introduction}
	An algebraic differential equation (ADE) is an equation of the form
	\begin{equation}\label{eq1}
		P(x,y(x),y'(x),\ldots,y^{(n)}(x))=0,
	\end{equation}
	where $P$ is a multivariate polynomial in $n+1$ variables over a field $\KK$. The positive integer $n$ is the order of the ADE, and the total degree of $P$ is simply called the degree of the ADE. We do not neglect $x$ because it may affect the resulting order. We define D-algebraic functions as solutions to ADEs. When the derivatives in \eqref{eq1} (including the $0$th derivative $y^{(0)}(x)=y(x)$) appear linearly, the ADE is called holonomic and its solutions are so-called D-finite functions. There are many computational and theoretical results related the D-finite functions in the literature. In recent years, attention has been drawn to interesting extensions of the class of D-finite functions, like DD-finite functions \cite{jimenez2018algorithmic} and $\delta_2$-finite functions \cite{TeguiaDelta2}.  
	
	A function is DD-finite if it satisfies a \textit{DD-finite equation}, i.e., a linear differential equation with D-finite function coefficients. These functions are D-algebraic \cite[Theorem 28]{jimenez2020some}, and share the same arithmetic properties with D-finite functions. We present an implementation for computing ADEs associated to DD-finite equations. This relies on a nonlinear algebra proof of the D-algebraic property of DD-finite functions. A $\delta_2$-finite function is a function that satisfies an ADE of degree at most $2$. The idea behind $\delta_2$-finite functions is to define normal forms and prove identities beyond holonomicity. See also \cite{TBguessing} for ongoing related work. The algorithmic concept may be generalized to higher degree ADEs. We present an implementation for \textit{looking for} ADEs of degree less than a given positive integer satisfied by a rational expression of D-algebraic functions. Note that this \textit{searching} approach does not rely on closure properties, and the underlying mathematics is more concerned with a smart search that increase computational efficiency by avoiding unnecessary computations. See \cite{hebisch2011extended} for similar interest.
	
	The main algorithms of our software relies on closure properties of D-algebraic functions. We implemented them from the ``constructive'' proofs given in \cite{RSB2023}. This is based on elimination with Gr\"obner bases. Our implementation is part of the package \texttt{NLDE} (\textbf{n}on\textbf{l}inear algebra and \textbf{d}ifferential \textbf{e}quations) written in the Maple language. We discuss comparisons with existing software in Section \ref{sec:disc}. The package can be used with any recent versions (from 2019 onward) of Maple. The source code and user-guide  instructions are given on the MathRepo \cite{mathrepo} repository at \url{https://mathrepo.mis.mpg.de/DAlgebraicFunctions/NLDEpackage}.
	\section{Arithmetic Operations}
	
	Let $\texttt{ADE}_1,\ldots,\texttt{ADE}_N$ be $N$ ADEs of orders $n_1,\ldots,n_N$, in the dependent variables $y_i(x)$, $i=1,\ldots, N$, respectively. Let $R$ be a multivariate rational function in $N+1$ variables. We mention that all computations are done over the field extension of the rationals defined by the polynomial coefficients, and we simply denote it by $\KK$. Suppose that the D-algebraic function $f_i\coloneqq f_i(x)$ is any solution of $\texttt{ADE}_i$. The aim of the procedures for arithmetic operations is to find a least-order ADE fulfilled by $F(x)\coloneqq R(x,f_1(x),\ldots,f_N(x))$. When $N\geq 2$, the underlying operation is said to be \textit{$N$-ary}. For $N=1$ we have a \textit{unary} operation. 
	
	\subsection{Computations Relying on Closure Properties}\label{sec:subsec1}
	
	We give insight for unary operations and refer the reader to \cite{RSB2023} and \cite{BT2023} for more details. We consider the ADE defined by \eqref{eq1}. We are looking for the least-order ADE satisfied by $R(x,f(x))$, where $R$ is a bivariate rational function, and $f$ is any solution of \eqref{eq1}. Let $m$ be the degree of $y^{(n)}$ in \eqref{eq1}. We can rewrite \eqref{eq1} as
	\begin{equation}\label{eq2}
		c_m\,{y^{(n)}}^m + P_1(x,y(x),y'(x),\ldots,y^{(n)}(x))=0,
	\end{equation}
	where $c_m\in S\coloneqq \KK[x,y(x),y'(x),\ldots,y^{(n-1)}(x)]$, $P_1\in S[y^{(n)}]$ such that $\deg_{y^{(n)}}(P_1)< m$. We make the ``change of variables'' $y^{(i)}\mapsto y_{i+1}$, for $i=0,\ldots,n-1$. Over the differential polynomial ring $\KK(x)\{y_1,\ldots,y_n,z\}$  with derivation $\frac{d}{dx}$, we consider the first $n$ derivatives of the $n+1$ differential polynomials
	\begin{equation}\label{eq3}
		z-R(x,y_1),\, c_m\,{y_{n-1}'}^m - P_1(x,y_1,\ldots,y_{n-1},y_{n-1}'),\, \ldots,\, y_1'-y_2. 
	\end{equation}
	The resulting set of polynomials can be viewed as a triangular set (see \cite[Section 4.2]{hubert2003notes}). The ideal defined by that triangular set has a non-trivial intersection with $\KK[x][z,z',\ldots,z^{(n)}]$. Therefore using Groebner bases elimination one can compute the ADE sought. Thus $R(x,f(x))$ is D-algebraic and we know how to find an ADE that it satisfies. For the output, we use the differential polynomial of lowest degree among those of the lowest possible order (here $\leq n$) among the generators of the elimination ideal.
	\begin{example}\label{ex:expl1} Suppose we want to find an ADE fulfilled by $\wp/(\wp+x)$ where $\wp$ is a Weirstrass elliptic function, i.e., it satisfies the ADE
		\begin{equation}\label{eq:wpeq}
			y'(x)^2 = 4y(x)^3-g_2\,y(x)-g_3.
		\end{equation}
	
		\vspace{-0.25cm}
	
		One proceeds as follows:
		
\begin{lstlisting}
> with(NLDE): #load the package from a library directory
> ADE:=diff(y(x),x)^2=4*y(x)^3-g2*y(x)-g3:
> unaryDalg(ADE,y(x),z=y/(x+y))
\end{lstlisting}
	
		\vspace{-0.5cm}
		
		\begin{dmath}\label{eq4}
			z \! \left(x \right)^{4} \left(4 x^{3}-\mathit{g2} x +\mathit{g3} +1\right)+z \! \left(x \right)^{3} \left(-4 x^{3}+3 \mathit{g2} x -4 \mathit{g3} -2\right)-2 z \! \left(x \right)^{2} \left(\frac{d}{d x}z \! \left(x \right)\right) x +z \! \left(x \right)^{2} \left(-3 \mathit{g2} x +6 \mathit{g3} +1\right)+2 x z \! \left(x \right) \left(\frac{d}{d x}z \! \left(x \right)\right)+z \! \left(x \right) \left(\mathit{g2} x -4 \mathit{g3} \right)+\left(\frac{d}{d x}z \! \left(x \right)\right)^{2} x^{2}+\mathit{g3} =0.
		\end{dmath}\QEDA
	\end{example}
	
	For $N$-ary operations, the idea is to construct a similar type of triangular set where the differential indeterminates $y_j$'s from \eqref{eq3} would play the role of the dependent variables of all the input ADEs. This establishes that $F(x)$ is D-algebraic and we know how to find a least-order (here $\leq n_1+\ldots+n_N$) ADE that it satisfies.
	
	\begin{example} It is well known that the ratio of two D-finite functions is generally not D-finite \cite{harris1985reciprocals}. The exponential generating function of Bernoulli polynomials $B_n(t)$ is given by $F(x)\coloneqq x\exp(t\,x)/(\exp(x)-1)$. The numerator and the denominator fulfill
		\begin{equation*}
			xy_1'(x) - (t\,x+1)\,y_1(x)=0, \, \text{ and }
			y_2'(x)-y_2(x) - 1 = 0,
		\end{equation*}
		
		\vspace{-0.15cm}
	
		respectively. An ADE satisfied by $F(x)$ is obtained by the following code:
		
\begin{lstlisting}
> ADE1:=x*diff(y1(x), x) - (t*x + 1)*y1(x) = 0:
> ADE2:=diff(y2(x), x)-y2(x) - 1=0:
> arithmeticDalg([ADE1,ADE2],[y1(x),y2(x)],z=y1/y2)
\end{lstlisting}
	
		\vspace{-0.5cm}
		
		\begin{dmath}\label{eq5}
			\left(-t^{2} x +t x -2 t +1\right) z \! \left(x \right)^{2}+\left(2 t x -x +2\right) \left(\frac{d}{d x}z \! \left(x \right)\right) z \! \left(x \right)-2 x \left(\frac{d}{d x}z \! \left(x \right)\right)^{2}+x \left(\frac{d^{2}}{d x^{2}}z \! \left(x \right)\right) z \! \left(x \right)=0.
		\end{dmath}\QEDA
	\end{example}
	\subsection{Computations Relying on Algorithmic Search (Linear Algebra)}
	
	Consider the ansatz
	
	\begin{equation}\label{eq6}
		\delta_k^d(F) + C_d \delta_k^{d-1}(F) + \cdots C_1\delta_k(F) + C_0,
	\end{equation}
	with unknown rational coefficients $C_i\coloneqq C_i(x), i=0,\ldots,d$; where $\delta_k^d$ is the $d$th monomial of degree at most $k$ w.r.t a chosen ranking for products of at most $k$ derivatives of $F$. For $j\in [1,N]\cap \NN $, using $\texttt{ADE}_j$, one can express higher derivatives of $y_j$ (representing $f_j$) in terms of its first $n_j$ derivatives. Substitution into \eqref{eq6} yields a rational expression whose numerator is a multivariate polynomial in  $\KK[y_j^{(i)},i=0,\ldots,n_j,j=1,\ldots,N]$. To have a valid equation for $F$, the coefficients of that polynomial must be zero. This yields a linear system in the $C_i$'s whose solutions lead to an ADE for $F$ after multiplication by the common denominator. 
	
	We implemented this approach as \texttt{Ansatz:-unaryDeltak} and \texttt{Ansatz:-arithmeticDeltak}, where \texttt{Ansatz} is a sub-package of \texttt{NLDE}.
	
	\begin{example}[\Cref{ex:expl1} continued] By default, \texttt{unaryDeltak} (and \texttt{arithmeticDeltak}) searches for an ADE of degree at most $k=2$. For \Cref{ex:expl1}, one finds a second-order quadratic ADE with the default setting. For $k=3$ a second-order cubic ADE is found. To recover \eqref{eq4}, one uses the following code:
		
\begin{lstlisting}
> Ansatz:-unaryDeltak(ADE,y(x),z=y/(x+y),degreeDE=4):
\end{lstlisting}
	
		We hide the output as it is identical to \eqref{eq4}.\QEDA
	\end{example}
	The syntax for \texttt{arithmeticDeltak} follows the same pattern with inputs given as for \texttt{arithmeticDalg}.
	
	\section{Other Operations}
	We start by presenting commands for computing ADEs for compositions, derivatives, and functional inverses of D-algebraic functions, and finish with the conversion of DD-finite equations into ADEs. 
	The algorithm for composition follows the same idea of \Cref{sec:subsec1}. For the derivatives and functional inverses, the output is quite explicitly constructed. Again, theoretical details are given in \cite{RSB2023}.
	\begin{example} Let us consider the operations in the duplication formula of the Weierstrass elliptic function:
		\begin{equation}\label{eq7}
			\wp(2x)-2\wp(x) = \frac{1}{4}\left(\frac{\wp''(x)}{\wp'(x)}\right)^2.
		\end{equation}
		For the left-hand side, we have the composition of $\wp$ and $2x$. We use the following code:
		
\begin{lstlisting}
> ADE:=diff(y1(x),x)^2=4*y1(x)^3-g2*y1(x)-g3:
> composeDalg([ADE,diff(y2(x),x)=2],[y1(x),y2(x)],y3(x))
\end{lstlisting}
	
		\vspace{-0.5cm}
		
		\begin{dmath}\label{eq9}
			-24 \mathit{y3} \! \left(x \right)^{2}+2 \mathit{g2} +\frac{d^{2}}{d x^{2}}\mathit{y3} \! \left(x \right)=0.
		\end{dmath}
		For the right-hand side, we use \texttt{diffDalg} to compute the derivatives.
		
\begin{lstlisting}
> diffDalg(ADE,y1(x),2)
\end{lstlisting}
	
		\vspace{-0.5cm}
		
		\begin{dmath}\label{eq10}
			16 \mathit{g2}^{5}+64 \mathit{g2}^{4} \text{\text{y1}} \! \left(x \right)+16 \mathit{g2}^{3} \text{y1} \! \left(x \right)^{2}-160 \mathit{g2}^{2} \text{y1} \! \left(x \right)^{3}-64 \mathit{g2} \text{y1} \! \left(x \right)^{4}+128 \text{y1} \! \left(x \right)^{5}-432 \mathit{g2}^{2} \mathit{g3}^{2}-1728 \mathit{g2} \,\mathit{g3}^{2} \text{y1} \! \left(x \right)-72 \mathit{g2} \mathit{g3} \left(\frac{d}{d x}\text{y1} \! \left(x \right)\right)^{2}-1728 \mathit{g3}^{2} \text{y1} \! \left(x \right)^{2}-144 \mathit{g3} \text{y1} \! \left(x \right) \left(\frac{d}{d x}\text{y1} \! \left(x \right)\right)^{2}-3 \left(\frac{d}{d x}\text{y1} \! \left(x \right)\right)^{4}=0.
		\end{dmath}
	
\begin{lstlisting}
> diffDalg(ADE,y1(x),1); #or simply diffDalg(ADE,y1(x))
\end{lstlisting}
	
		\vspace{-0.5cm}
		
		\begin{dmath}\label{eq11}
			-1728 \mathit{\text{y1}} \! \left(x \right)^{4}+64 \mathit{g2}^{3}-192 \mathit{g2} \left(\frac{d}{d x}\mathit{\text{y1}} \! \left(x \right)\right)^{2}-3456 \mathit{g3} \mathit{\text{y1}} \! \left(x \right)^{2}+128 \left(\frac{d}{d x}\mathit{\text{y1}} \! \left(x \right)\right)^{3}-1728 \mathit{g3}^{2}=0.
		\end{dmath}
		To verify the duplication formula one computes a forth-order ADE for both sides of the identities.
		
		As a last computation in this series, let us find an ADE for functional inverses of $\wp$.
		
\begin{lstlisting}
> invDalg(ADE,y1(x),z(x))
\end{lstlisting}
	
		\vspace{-0.5cm}
		
		\begin{dmath}\label{(1)}
			1+\left(-4 x^{3}+\mathit{g2} x +\mathit{g3} \right) \left(\frac{d}{d x}z \! \left(x \right)\right)^{2}=0.
		\end{dmath}
		
		\QEDA
	\end{example}

	We end by computing an ADE for Mathieu functions. These are solutions of the differential equation
	\begin{equation}\label{eq12}
		y^{''}(x) + (a-2\,q\,\cos(2\,x))\,y(x)=0,
	\end{equation}
	where $q$ and $a$ are some parameters. We use our command \texttt{DDfiniteToDalg}. 
	\begin{example}
		We first compute a holonomic ODE satisfied by $\cos(2\,x)$ and replace it in \eqref{eq12} by the name of the dependent variable in that holonomic ODE.
		
\begin{lstlisting}
> cosDE:=DEtools:-FindODE(cos(2*x),C(x)):
> ADE:=diff(y(x),x,x)+(a-2*q*C)*y(x)=0:
> DDfiniteToDalg(ADE,y(x),[cosDE],[C(x)])
\end{lstlisting}
	
		\vspace{-0.5cm}
		
		\begin{dmath}\label{eq:eq13}
			4 a y \! \left(x \right)^{3}+4 y \! \left(x \right)^{2} \left(\frac{d^{2}}{d x^{2}}y \! \left(x \right)\right)+y \! \left(x \right)^{2} \left(\frac{d^{4}}{d x^{4}}y \! \left(x \right)\right)-2 \left(\frac{d^{3}}{d x^{3}}y \! \left(x \right)\right) y \! \left(x \right) \left(\frac{d}{d x}y \! \left(x \right)\right)-\left(\frac{d^{2}}{d x^{2}}y \! \left(x \right)\right)^{2} y \! \left(x \right)+2 \left(\frac{d}{d x}y \! \left(x \right)\right)^{2} \left(\frac{d^{2}}{d x^{2}}y \! \left(x \right)\right)=0.
		\end{dmath}
		In general, the third and the forth arguments in the input are lists of the corresponding holonomic ODEs and their dependent variables, respectively.\QEDA
	\end{example}

\section{Discussions}\label{sec:disc}

There are existing software packages that can be used to perform some of our operations. For instance, the \texttt{find_ioequations} command (see \cite{ovchinnikov2020computing}) of the Julia \cite{Julia} package \texttt{StructuralIdentifiability} may derive some ADEs with constant coefficients only; moreover, that command requires a user-defined dynamical model, which is not needed in our algorithms. More interesting comparable software packages include the \texttt{RosendfeldGroebner} command of the Maple \texttt{DifferentialAlgebra} package based on the library at \cite{DifferentialAlgebra}, and the \texttt{ThomasDecomposition} command of the Maple \texttt{DifferentialThomas} package (see \cite[Section 2.2]{robertz2014formal}). However, the main difference with our algorithm is that \texttt{RosendfeldGroebner} and \texttt{ThomasDecomposition} return an element of the radical differential ideal, whereas our algorithm returns an element of the differential ideal. In many cases, when the highest order terms of the input ADEs appear only linearly, \texttt{RosendfeldGroebner} seems more efficient. On the other hand, with input ADEs whose highest order terms do not appear linearly, our algorithm generally outperforms \texttt{RosendfeldGroebner} and \texttt{ThomasDecomposition}. For instance, for \cite[Example 5]{BT2023}, \texttt{RosendfeldGroebner} runs out of memory, and \texttt{ThomasDecomposition} keeps running even after an hour of computation. Therefore, on its own, our implementation provides a relevant tool for differential elimination. We are thinking of combining our implementation with \texttt{RosendfeldGroebner} and/or \texttt{ThomasDecomposition} to obtain a satisfactory efficiency in all situations. This future improvement may be particularly important for the command \texttt{arithmeticMDalg} of \texttt{NLDE:-MDalg}, which is concerned with multivariate D-algebraic functions (see \cite{BT2023}).

\section*{Acknowledgment} We thank the anonymous referees for initiating Section \ref{sec:disc} and pointing out \cite{DifferentialAlgebra}.


\begin{thebibliography}{10}
	\providecommand{\url}[1]{\texttt{#1}}
	\providecommand{\urlprefix}{URL }
	\providecommand{\doi}[1]{https://doi.org/#1}
	
	\bibitem{Julia}
	Bezanson, J., Edelman, A., Karpinski, S., Shah, V.B.: Julia: A fresh approach
	to numerical computing. SIAM {R}eview  \textbf{59}(1),  65--98 (2017).
	\doi{10.1137/141000671}, \url{https://epubs.siam.org/doi/10.1137/141000671}
	
	\bibitem{DifferentialAlgebra}
	Boulier, F.: {DifferentialAlgebra project: A {C} library for differential
		elimination}. Available at
	\url{https://codeberg.org/francois.boulier/DifferentialAlgebra}
	
	\bibitem{mathrepo}
	Fevola, C., G\"{o}rgen, C.: The mathematical research-data repository
	{M}ath{R}epo. {C}omputeralgebra {R}undbrief  \textbf{70},  16--20 (2022)
	
	\bibitem{harris1985reciprocals}
	Harris~Jr, W.A., Sibuya, Y.: The reciprocals of solutions of linear ordinary
	differential equations. Advances in Mathematics  \textbf{58}(2),  119--132
	(1985)
	
	\bibitem{hebisch2011extended}
	Hebisch, W., Rubey, M.: Extended rate, more {GFUN}. Journal of Symbolic
	Computation  \textbf{46}(8),  889--903 (2011)
	
	\bibitem{hubert2003notes}
	Hubert, E.: Notes on triangular sets and triangulation-decomposition algorithms
	i: Polynomial systems. In: Symbolic and Numerical Scientific Computation:
	Second International Conference, SNSC 2001, Hagenberg, Austria, September
	12--14, 2001. Revised Papers. pp. 1--39. Springer (2003)
	
	\bibitem{jimenez2018algorithmic}
	Jim{\'e}nez-Pastor, A., Pillwein, V.: Algorithmic arithmetics with dd-finite
	functions. In: Proceedings of the 2018 ACM International Symposium on
	Symbolic and Algebraic Computation. pp. 231--237 (2018)
	
	\bibitem{jimenez2020some}
	Jim{\'e}nez-Pastor, A., Pillwein, V., Singer, M.F.: Some structural results on
	dn-finite functions. Advances in Applied Mathematics  \textbf{117},  102027
	(2020)
	
	\bibitem{RSB2023}
	Manssour Ait~El, R., Sattelberger, A.L., Teguia~Tabuguia, B.: D-algebraic
	functions. arXiv preprint arXiv:2301.02512  (2023)
	
	\bibitem{ovchinnikov2020computing}
	Ovchinnikov, A., Pillay, A., Pogudin, G., Scanlon, T.: Computing all
	identifiable functions for {ODE} models. arXiv preprint arXiv:2004.07774
	(2020)
	
	\bibitem{robertz2014formal}
	Robertz, D.: Formal algorithmic elimination for PDEs, vol.~2121. Springer
	(2014)
	
	\bibitem{TBguessing}
	Teguia~Tabuguia, B.: Guessing with quadratic differential equations (2022),
	{S}oftware {D}emo at {ISSAC'22}.
	
	\bibitem{BT2023}
	Teguia~Tabuguia, B.: Arithmetic of {D}-algebraic functions. arXiv preprint
	arXiv:2305.00702  (2023)
	
	\bibitem{TeguiaDelta2}
	Teguia~Tabuguia, B., Koepf, W.: On the representation of non-holonomic
	univariate power series. Maple Trans.  \textbf{2}(1) (2022), article 14315
	
\end{thebibliography}
\end{document}